\documentclass[12pt]{iopart}
\usepackage{graphicx}

\usepackage{iopams}
\usepackage{xcolor}

\usepackage{citesort}
\begin{document}

\title[QCD at finite chemical potential in and out-of equilibrium]
{QCD at finite chemical potential in and out-of equilibrium}

\author{Olga Soloveva$^1$, Pierre Moreau$^2$, Elena Bratkovskaya$^{3,1,4}$ }

\address{}
\address{$^1$ Institute for Theoretical Physics,
University of Frankfurt, Frankfurt, Germany}
\address{$^2$ Department of Physics, Duke University, Durham, NC 27708, USA}
\address{$^3$ GSI, Helmholtzzentrum f\"{u}r Schwerionenforschung GmbH, Darmstadt, Germany}
\address{$^4$ Helmholtz Research Academy Hessen for FAIR (HFHF), GSI Helmholtz Center for Heavy Ion Physics, Campus Frankfurt, 60438 Frankfurt, Germany}

\ead{E.Bratkovskaya@gsi.de}

\vspace{10pt}
\begin{indented}
\item[]March 2021
\end{indented}

\begin{abstract}
We review the transport properties of the strongly interacting quark-gluon plasma (QGP)
created in heavy-ion collisions at ultrarelativistic energies, i.e.
out-of equilibrium, and compare them to the equilibrium properties.
The description of the strongly interacting (non-perturbative) QGP in equilibrium
is based on the effective propagators and couplings from the Dynamical QuasiParticle
Model (DQPM) that is matched to reproduce the equation-of-state of the partonic system
above the deconfinement temperature $T_c$ from lattice QCD.
We study the transport coefficients such as the ratio of shear viscosity and bulk viscosity
over entropy density, diffusion coefficients, electric conductivity etc. versus temperature and baryon chemical potential. Based on a microscopic transport description of heavy-ion collisions
we, furthermore, discuss which observables are sensitive to the QGP formation and its properties.
\end{abstract}

\vspace{2pc}
\submitto{\PS}
\maketitle

\section{Introduction}
%\bibliography{Biblio-brat}

%\bibliography{iopart-num}

%\end{document}
     Numerous achievements of heavy-ion collision (HIC) experiments have dramatically changed the theoretical understanding of the QCD matter properties, especially the deconfined QCD matter created in the central interaction volume at relativistic energies. Initially the deconfined QCD matter - or the Quark-Gluon-Plasma (QGP) - has been considered as a weakly interacting system of massless partons (quarks and gluons) which might be described by perturbative QCD (pQCD). Measurements of the anisotropic particle flow and jet quenching  at the Relativistic Heavy-Ion Collider (RHIC) revealed that in the vicinity of the quark-hadron phase transition the deconfined state of QCD matter (at high temperature and partonic density) behaves as a nearly perfect relativistic fluid with the lowest value of the specific shear viscosity among the known fluids \cite{Arsene:2004fa,Adams:2005dq,Shuryak:1978ij,Gyulassy:2004zy}. Actual and future heavy-ion collision experiments will probe the QCD phase diagram at nonzero baryon density (or baryon chemical potential). At ultra-relativistic energies at the Large Hadron Collider (LHC) or the Relativistic Heavy-Ion Collider (RHIC) the quark-gluon plasma is formed with nearly zero baryon chemical potential and the phase transition from the partons to the  hadrons is a rapid crossover.  In order to quantify experimental findings one can employ  transport or hydrodynamic simulations that successfully describe the differential distributions of produced particles in these HICs.

    While there are a plethora of studies about the QCD medium in the confined hadronic phase the future facilities will explore the QCD medium at high net baryon density or baryon chemical potential. The QCD phase diagram can be understood from the thermodynamic point of view in terms of the temperature $T$ and baryon chemical potential $\mu_B$, where the most unexplored region is located at moderate temperatures and relatively high $\mu_B$. This region is of particular  interest in the Beam Energy Scan programs at RHIC \cite{Odyniec:2019kfh} as well as the future experimental program of FAIR (Facility for Antiproton and Ion Research) \cite{Senger:2020fvj} at GSI and the NICA(Nuclotron-based Ion Collider fAcility) facility at JINR \cite{Sissakian:2009zza}. In the region of vanishing baryon chemical potential one can apply lattice QCD (lQCD) techniques to describe the macroscopic properties of the QCD medium such as the equation of state; moreover one can extract transport coefficients of the QCD medium at $\mu_B =0$. A primary difficulty in dealing with non-zero baryon chemical potentials is the fermion sign problem for QCD (the fermion determinant is not positive definite), which makes the current lattice QCD methods inapplicable in the low $T$ and finite $\mu_B$ region.

An alternative approach, which can describe microscopic properties of the deconfined QCD medium in a wide range of baryon chemical potentials, is to use effective models on the basis of 'resummed' propagators and couplings. In this work we review the results of the transport properties and the evolution of the QCD medium at finite chemical potential $\mu_B $ in equilibrium while for the description of the deconfined phase - near and out equilibrium - effective models and the PHSD transport approach are applied. Considering the transport coefficients and the EoS of the QGP phase we compare our results with the various results from the literature at vanishing chemical potential.

To quantify the strongly-interacting liquid one can apply hydrodynamic descriptions of the system. However, in order to perform hydrodynamical simulations of the time evolution of the quark-gluon matter at finite baryon chemical potential, one needs to estimate or know the transport coefficients of the matter in this region. The evaluation of the transport coefficients at finite $\mu_B$ depends on the underlying microscopic theory which describes the interaction between quarks and gluons, but we face a fundamental problem to construct and evaluate such a theory at finite $T$ and $\mu_B$ from first principles.

\section{Microscopic properties of the QGP at finite chemical potential}

For the full description of the QGP dynamics one needs to estimate first the microscopic properties of the relevant degrees of freedom  such as the effective masses (and widths) of the partonic propagators as well as elastic cross-sections for the different partons.
In the region of finite $T$ and moderate baryon chemical potential $\mu_B$ one can rely on the estimates from effective models.
We here consider essential features of the QGP medium in terms of strongly interacting quarks and gluons as given by the dynamical quasi-particle model (DQPM) \cite{Peshier:2005pp,Cassing:2007nb,Cassing:2007yg}. The DQPM  reproduces the equation of state of the partonic system  above  the  deconfinement  temperature $T_c$ from  lattice  Quantum  Chromodynamics (QCD) and predicts reasonable estimates for the QGP transport coefficients, which - as we will see later - are in agreement with the lQCD results available at $\mu_B = 0 $.
In the DQPM the quasi-particles are characterized by dressed propagators with complex self-energies, where the real part of the self-energies is related to  dynamically  generated  thermal masses, while the imaginary part provides information about the lifetime and reaction rates of the partons (interaction widths),
\begin{equation}
\label{propdqpm} G^{R} (\omega, {\bf p}) = \frac{1}{\omega^2 - {\bf
p}^2 - M^2 + 2 i \gamma \omega} ,
\end{equation}
~\\
using $\omega=p_0$ for energy, while $ M$ and $\gamma$ are the thermal mass and width of a parton.
In the DQPM the spectral functions of the quasiparticles $\rho_j$ ($j=q, {\bar q}, g$) or imaginary parts of the propagator $\rho_j = - 2 \mathrm{Im}( G^R_j)$  are no
longer $\delta$-functions in the invariant mass squared but given by
\begin{equation}
\rho_{j}(\omega,{\bf p}) =  \frac{4\omega\gamma_j}{\left( \omega^2 - \mathbf{p}^2 - M^2_j \right)^2 + 4\gamma^2_j \omega^2}
\label{spectral_function}
\end{equation}
~\\
separately for quarks, antiquarks and gluons ($j = q,\bar q,g$).

\begin{figure}[!ht]
 \centerline{\includegraphics[scale=0.8]{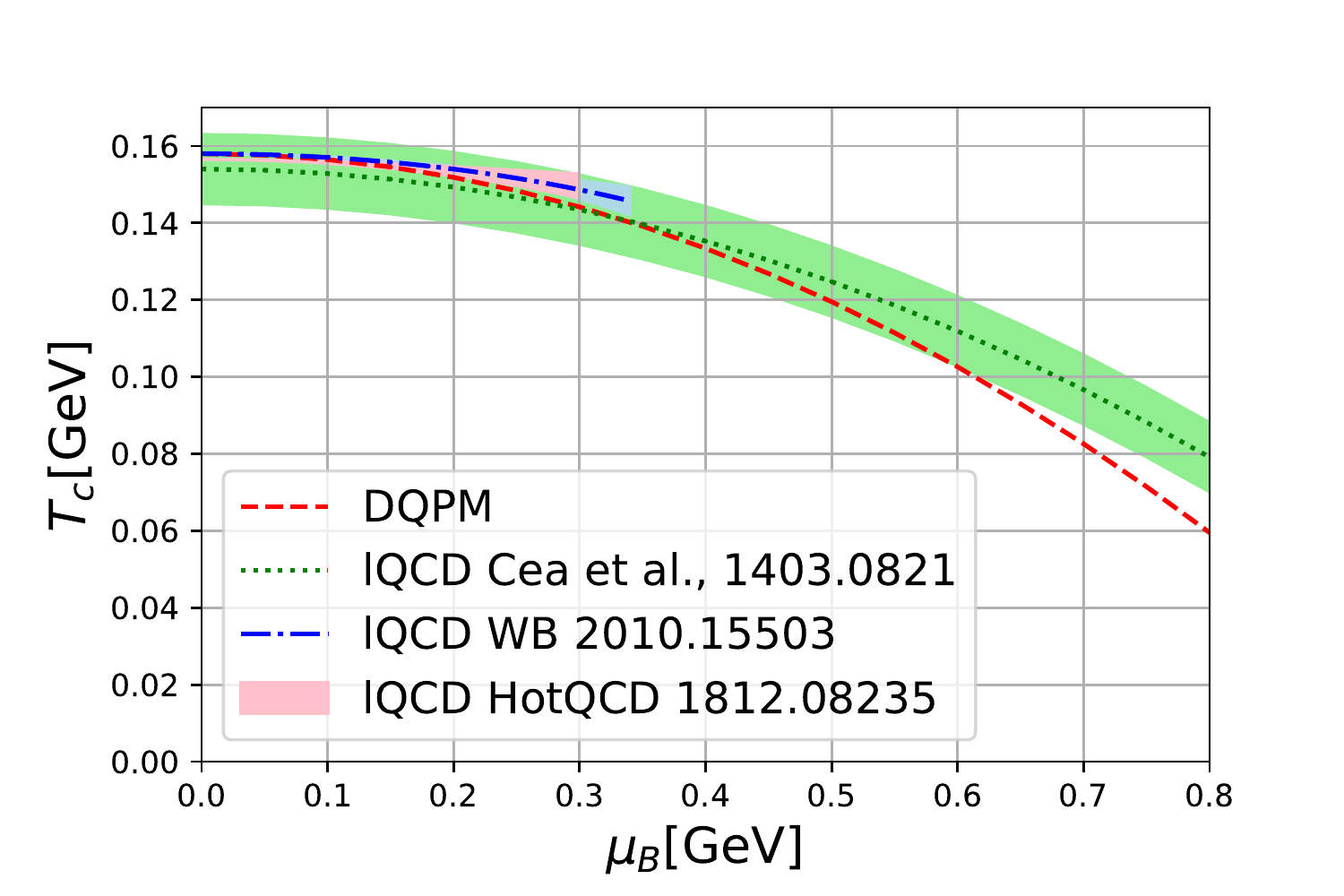} }
\caption{(Pseudo-)critical line $T_c(\mu_B)$. The red dashed line corresponds to the DQPM phase boundary. Lines with colored areas represent lQCD estimates of $T_c(\mu_B)$ for QCD with $N_f= 2 + 1$: (green area) \cite{Cea:2014xva}, (blue area) \cite{Guenther:2020vqg,Borsanyi:2020fev}, (red area) \cite{Bazavov:2018mes}.}
\label{fig:Tcmu}
\end{figure}

All the microscopic properties as well as transport coefficients of the QCD matter are sensitive to the underlying coupling of the matter, which can be estimated within the DQPM using the entropy density from lQCD at $\mu_B =0$. Although  the coupling constant $g^2$ in general depends on temperature and baryon chemical potential, we start with the determination of $g^2(T, \mu_B = 0 )$. The temperature dependence is parameterized via the entropy density $s(T,\mu_\mathrm{B} = 0)$ from lattice QCD from Refs. \cite{Borsanyi:2012cr,Borsanyi:2013bia} in the following way:
	\begin{equation}
	g^2(T,\mu_\mathrm{B} = 0) = d \Big( \left(s(T,0)/s^\mathrm{QCD}_{\mathrm{SB}} \right)^e -1 \Big)^f,
	\label{coupling_DQPM}
	\end{equation}
	with the Stefan-Boltzmann entropy density $s_{\mathrm{SB}}^{\mathrm{QCD}} = 19/9\ \pi^2 T^3$ and the dimensionless parameters $d = 169.934$, $e = -0.178434$ and $f = 1.14631$.
	In order to obtain the coupling at finite baryon chemical potential $\mu_\mathrm{B}$, we use the 'scaling hypothesis' introduced in \cite{Cassing:2008nn}, which assumes that $g^2$ is a function of the ratio of the effective temperature $T^* = \sqrt{(T^2+\mu^2_q/\pi^2)^{1/2}}$ ( where quark chemical potential is defined as follows $\mu_q=\mu_u=\mu_s=\mu_B/3$ ) and the $\mu_\mathrm{B}$-dependent critical temperature $T_c(\mu_\mathrm{B})$ as \cite{Berrehrah:2016vzw}
	\begin{equation}
	g^2(T/T_c,\mu_\mathrm{B}) = g^2\left(\frac{T^*}{T_c(\mu_\mathrm{B})},\mu_\mathrm{B} =0 \right),
	\label{coupling}
	\end{equation}
	with  $T_c(\mu_B) = T_c (1-\alpha \mu_B^2)^{1/2}$ being the (pseudo)critical line of the DQPM (see Fig. \ref{fig:Tcmu}), where $T_c$ is the critical temperature at vanishing chemical potential ($\approx 0.158$ GeV) and $\alpha = 0.974$ GeV$^{-2}$.

One can define the real part of the partonic self-energies by the dynamical quasi-particle masses (for gluons and quarks) which are assumed to be given by the HTL thermal masses in the asymptotic high-momentum regime by \cite{Bellac:2011kqa,Linnyk:2015rco}
	\begin{equation}
	 m^2_{g}(T,\mu_\mathrm{B})=\frac{g^2(T,\mu_\mathrm{B})}{6}\left(\left(N_{c}+\frac{N_{f}}{2}\right)T^2
	+\frac{N_c}{2}\sum_{q}\frac{\mu^{2}_{q}}{\pi^2}\right)
		\end{equation}
		\begin{equation}
	  m^2_{q(\bar q)}(T,\mu_\mathrm{B})=\frac{N^{2}_{c}-1}{8N_{c}}g^2(T,\mu_\mathrm{B})\left(T^2+ \frac{\mu^{2}_{q}}{\pi^2}\right)
	  \label{mass},
	\end{equation}
	where $N_{c}=3$ is the number of colors, while $N_{f} =3$ denotes the number of flavors. The strange quark has a larger bare mass which needs to be considered in its dynamical mass. Empirically we find $m_s(T,\mu_\mathrm{B})= m_{u/d}(T,\mu_\mathrm{B})+ \Delta m$ and $\Delta m \approx$
	30 MeV.	Furthermore, the quasi-particles in the DQPM have finite widths, which are adopted in the form \cite{Berrehrah:2016vzw,Linnyk:2015rco}
	\begin{equation}
		\gamma_{i}(T,\mu_\mathrm{B}) = \frac{1}{3} C_i \frac{g^2(T,\mu_\mathrm{B})T}{8\pi}\ln\left(\frac{2c_m}{g^2(T,\mu_\mathrm{B})}+1\right),
	\end{equation}
where we use the QCD color factors for quarks, $C_q = C_F = \frac{N_c^2 - 1}{2 N_c} = 4/3$, and for gluons $C_g = C_A = N_c = 3$. Further, we fixed the parameter $c_m = 14.4$, which is related to a magnetic cut-off. We assume that the width of the strange quark is the same as that for the light ($u,d$) quarks.
Once  the quasi-particle properties (or propagators) are fixed as described above, one can
evaluate the entropy density $s(T,\mu_\mathrm{B})$, the pressure $P(T,\mu_\mathrm{B})$ and energy density $\epsilon(T,\mu_\mathrm{B})$ in a straight forward manner by starting with the entropy density $s^{\mathrm{dqp}}$ and number density $n^{\mathrm{dqp}}$ in the propagator representation from Baym \cite{Baym,Blaizot:2000fc} and then identifying $s = s^{\mathrm{dqp}}$ and $n_B = n^{\mathrm{dqp}}/3$ \cite{Soloveva:2019xph}. The isotropic pressure $P_0$ can then be obtained by using the Maxwell relation of a grand canonical ensemble:	
	\begin{equation}
	P(T,\mu_\mathrm{B}) =
	 P(T_0,0)  + \int\limits_{T_{0}}^{T} s(T',0)\ dT'+ \int\limits_{0}^{\mu_\mathrm{B}} n_B(T,\mu_\mathrm{B}')\ d\mu_\mathrm{B}' \label{pressure} ,
	\end{equation}
	where the lower bound in temperature is chosen between $0.1 < T_0 < 0.15$ GeV.
	The energy density $\epsilon$ then follows from the Euler relation
	\begin{equation}
	\label{eps} \epsilon = T s - P +\mu_\mathrm{B} n_\mathrm{B} .
	\end{equation}

 The differential cross section for a binary process of on-shell particles ($ i + j \rightarrow c + d$) in the center-of-momentum frame (CM), where the momenta of the colliding particles obey $\mathbf{k}_i + \mathbf{k^\prime}_j = \mathbf{p}_c + \mathbf{p^\prime}_d = \mathbf{P} = 0$ and $k^0_i + {k^\prime}^0_j = \sqrt{s} = p^0_i + {p^\prime}^0_j$, is given by	
	\begin{equation}
	\mathrm{d}\sigma^{on}_{ij \rightarrow cd}(\sqrt{s},\Omega) = \frac{1}{64 \pi^2 s} \frac{p_{\mathrm{out}}}{p_\mathrm{in}} |\bar{\mathcal{M}}|^2 \mathrm{d}\Omega,
	\label{dsigma_on_CM}
	\end{equation}
	where  $\bar{\mathcal{M}}^2 \equiv \bar{\mathcal{M}}^2(k_{i_1} \dots k_{i_n} \rightarrow p_{j_1} \dots p_{j_m})$ is the invariant matrix element squared averaged over the color and spin of the incoming particles and summed over those of the final particles. In (\ref{dsigma_on_CM}) $\mathrm{d}\Omega$ is the differential solid angle corresponding to one of the final particles. The momenta of the initial ($p_\mathrm{in}$) and final particles ($p_\mathrm{out}$) in the CM frame are found to be,
	\begin{equation}
	p_{i} = \frac{\sqrt{\left(s-(m_{i} + m^\prime_{i})^2\right)\left(s-(m_{i}-m^\prime_{i})^2\right)}}{2\sqrt{s}} ,
	\end{equation}
	~\\
	where $i= \mathrm{in}/\mathrm{out}$, $m_{i}$ and $m^\prime_{i}$ being the masses of the colliding partons.
	
	The total cross section is obtained via:
	\begin{equation}
\sigma^{ij \rightarrow cd}_{\mathrm{tot}}(\sqrt{s}) = \frac{1}{32 \pi s} \frac{p_{out}}{p_{in}} \gamma_{ij} \int_{-1}^{1} d \cos(\theta)  \ |\bar{\mathcal{M}}|^2 ,
\label{sigma_on_CM}
\end{equation}
where $\theta$ is the final polar angle of one of the final particles in the CM frame, and
$\gamma_{ij} = 1 - \frac{1}{2}\delta_{ij}$ is the symmetry factor.
	
%----------------------------cross-sections DQPM ------------------------------------------------------
\begin{figure}[!ht]
 \includegraphics[scale=0.35]{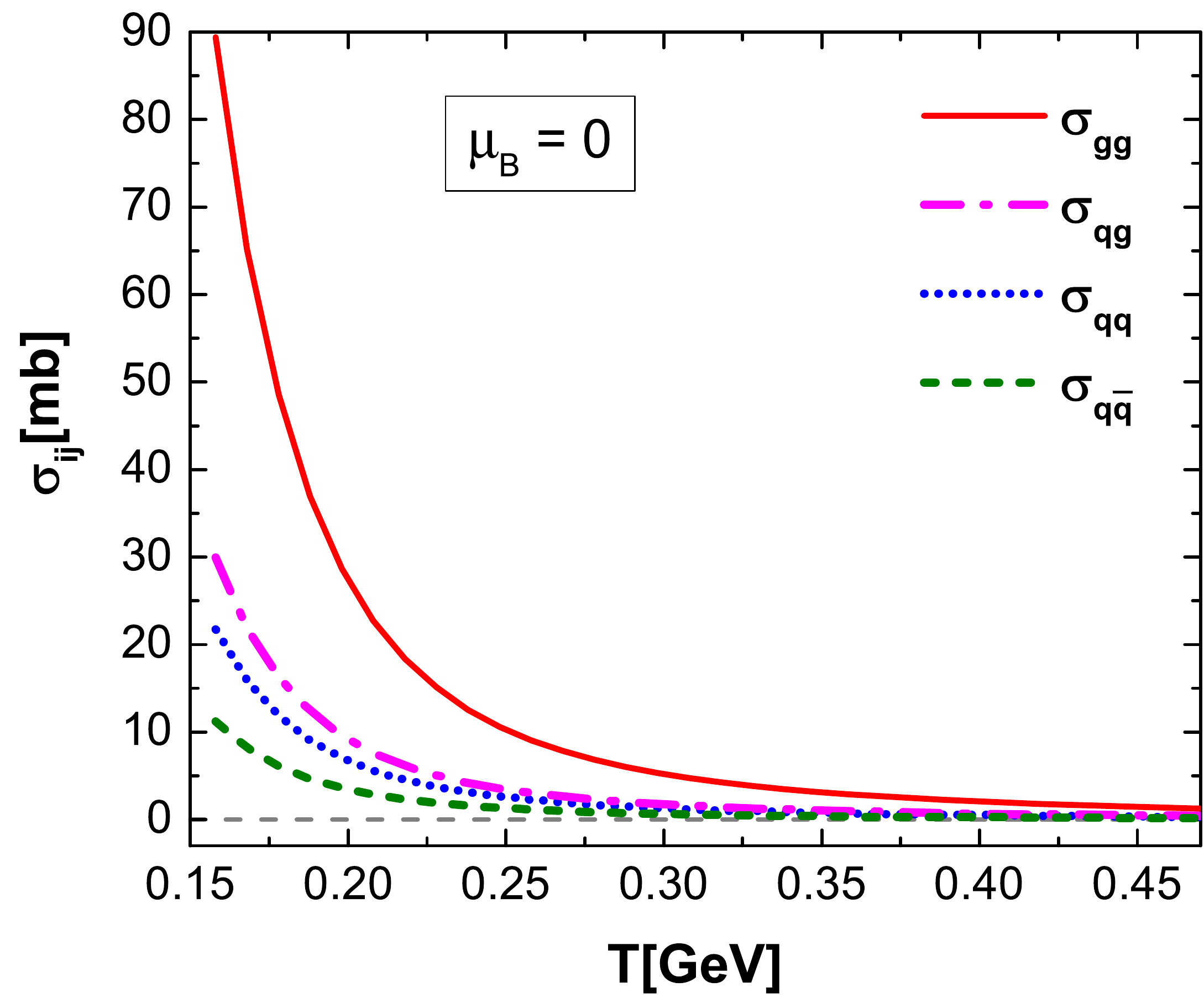}
 \includegraphics[scale=0.35]{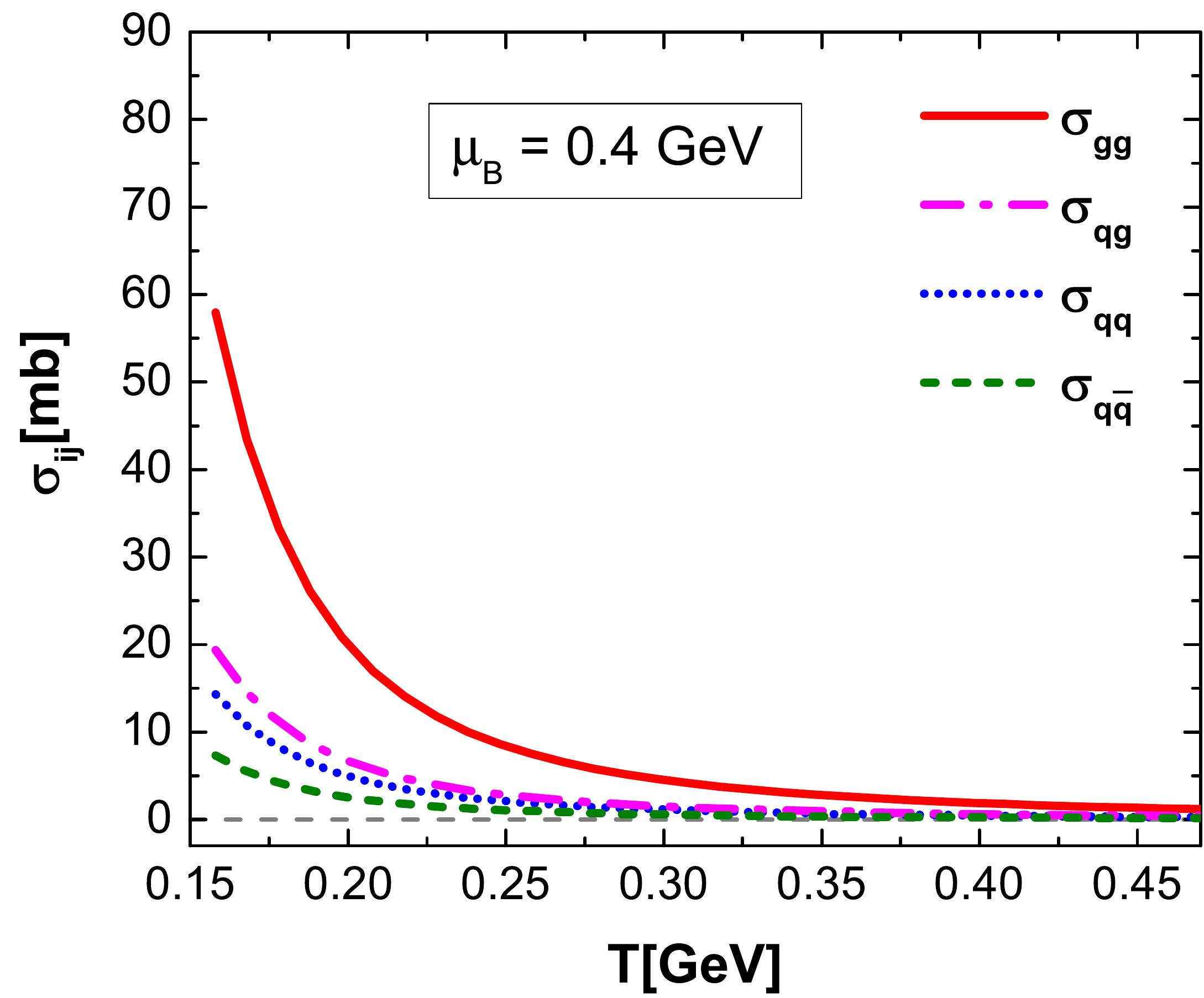}
\caption{Total DQPM parton elastic cross sections  as a function of temperature at $\mu_B=0$ (l.h.s.) and at $\mu_B=0.4$ GeV (r.h.s.) from Ref.\ \cite{Moreau:2019vhw}.  }
\label{fig:sigma}
\end{figure}
%-------------------------------------------------------------------------------------------------------
	
The DQPM total cross-sections for different channels are shown in Fig. \ref{fig:sigma} as a function
of temperature for $\mu_B =0$ and $\mu_B =0.4$ GeV. The total cross sections depend on temperature
as $\sigma_{ij} \sim 1/T^3$ or $\sim 1/T^4$, which is governed by the $T$-dependence of the DQPM
running coupling. The difference in the $T$-dependence for different channels arises from the
combinations of $s -, t-, u-$ channels: for $q-q$, $q-\bar{q}$ and $q-g$ scatterings $\sigma_{ij}
\sim 1/T^3$, while for the $g-g$ channel the terms $1/T^3$, $1/T^4$ have equivalent contributions to
the total cross-section $\sigma_{ij} \sim c_3/T^3+c_4/T^4$, where $c_3,c_4$ depend on $\sqrt{s},
\mu_B$. The cross-sections decrease with increasing chemical potential as expected from the $\mu_B$
dependence of the DQPM coupling. This trend is in agreement with the pQCD predictions for  high
temperatures and chemical potential, where the QGP can be described approximately by a free gas of
quarks and gluons.
We have found that the collisional widths from the DQPM and those evaluated from
the interaction rates - by calculating the partonic differential cross-sections
as a function of $T$ and $\mu_B$ for the leading tree-level diagrams -
are in  reasonable agreement. Accordingly, the quasi-particle limit holds sufficiently well \cite{Moreau:2019vhw}.

For the  scatterings of the on-shell partons (i.e. the energies of the particles are taken to be $E^2 = \mathbf{p}^2 + M^2$ with $M$ being the pole mass) the interaction rate $\Gamma^{on}$ is obtained as follows,
\begin{equation}
\Gamma^{on}_i  (p_i, T,\mu_q) = \sum_{j=q,\bar{q},g} \int \frac{d^3p_j}{(2\pi)^3}\ d_j\ f_j\ v_{rel} \int d\sigma^{on}_{ij \rightarrow cd}\ (1\pm f_c) (1\pm f_d) ,
\label{Gamma_on}
\end{equation}
where $v_{rel} =\frac{\sqrt{(p_i \cdot p_j)^2-m_i^2 m_j^2}}{E_i E_j}$ denotes the relative velocity in the c.m. frame, $d_j$ is the degeneracy factor for spin and color (for quarks $d_q = 2 \times N_c$ and for gluons $d_g =2 \times (N_c^2-1)$), using the shorthand notation $f_j = f_j(E_j,T,\mu_q)$ for the distribution functions. In Eq. (\ref{Gamma_on}) and in all the following sections, the notation $\sum_{j=q,\bar{q},g}$ includes the contribution from all possible partons which in our case are the gluons and the (anti-)\-quarks of three different flavors ($u,d,s$). The parton interaction rate $\Gamma^{on}_i  (p_i, T,\mu_q)$ can be used for the estimation of the parton relaxation time, which is the key ingredient for transport coefficients from kinetic theory.

\section{Transport coefficients of the QCD at finite chemical potential}
We start with the transport coefficients at vanishing baryon chemical potential.
We compute the ratio of the shear and bulk viscosities to the
entropy density, i.e., $\eta/s$ and $\zeta/s$, the electric conductivity
$\sigma_0/T$, as well as the baryon-diffusion coefficient $\kappa_B/T^2$
and found that the ratios $\eta/s$ and $\zeta/s$ as well as $\sigma_0/T$
are in accord with the available results from lQCD at $\mu_B=0$.

The most popular transport coefficients - shear and bulk viscosities - have been evaluated within various models in the hadronic and the partonic phase. The shear viscosity reflects the strength of the interaction inside the QGP medium; it can be related to the parton interaction rates, which is a challenge to evaluate on the basis of first principles. Various theoretical models show that the temperature dependence of the QCD specific shear viscosity is qualitatively different in the confined and deconfined phases. Starting from the hadronic phase below the pseudo-critical temperature $T<T_c$, $\eta/s$ monotonically decreases with $T$ because the system is dominated by pions with weaker interactions at lower $T$. For the partonic phase above the pseudo-critical temperature $T>T_c$, $\eta/s$  increases with temperature because the interaction attenuates at high T. Approaching the phase transition from hadronic to partonic phase at vanishing chemical potential the specific shear viscosity shows a dip followed by an increase with temperature. A similar temperature dependence of the specific shear viscosity $\eta/s$ is also seen for other fluids such as $H_2O$ and $He$.

While first only the shear viscosity was considered in the hydrodynamical simulations it was found later that the bulk viscosity of the QGP should be non-zero, at least in the vicinity of the phase transition \cite{Ryu:2015vwa}. The bulk viscosity reduces the speed of the radial expansion and thus influences the mean momentum of produced particles.
It is known that the bulk viscosity is identically zero in conformal fluids and it is expected that the QCD medium approaches a conformal behaviour in the high-energy or temperature limit. Nevertheless, the lQCD results on the enhanced trace anomaly close to $T_c$ have shown that it is probably not the case for the deconfined QCD medium in the vicinity of the phase transition.

One way to evaluate the viscosity coefficients of QCD matter is the Kubo formalism \cite{Kubo:1957mj,Aarts:2002cc}, which was used to calculate the viscosities for a previous version of the DQPM within the PHSD transport approach in a box with periodic boundary conditions (cf. Ref. \cite{Ozvenchuk:2012kh}). The method becomes notoriously difficult for the partonic phase. A more simple way to estimate the transport coefficient is to imply the effective Boltzmann equation in the relaxation-time approximation (RTA) \cite{Chakraborty:2010fr}.

\begin{figure}[!ht]
 \includegraphics[scale=0.39]{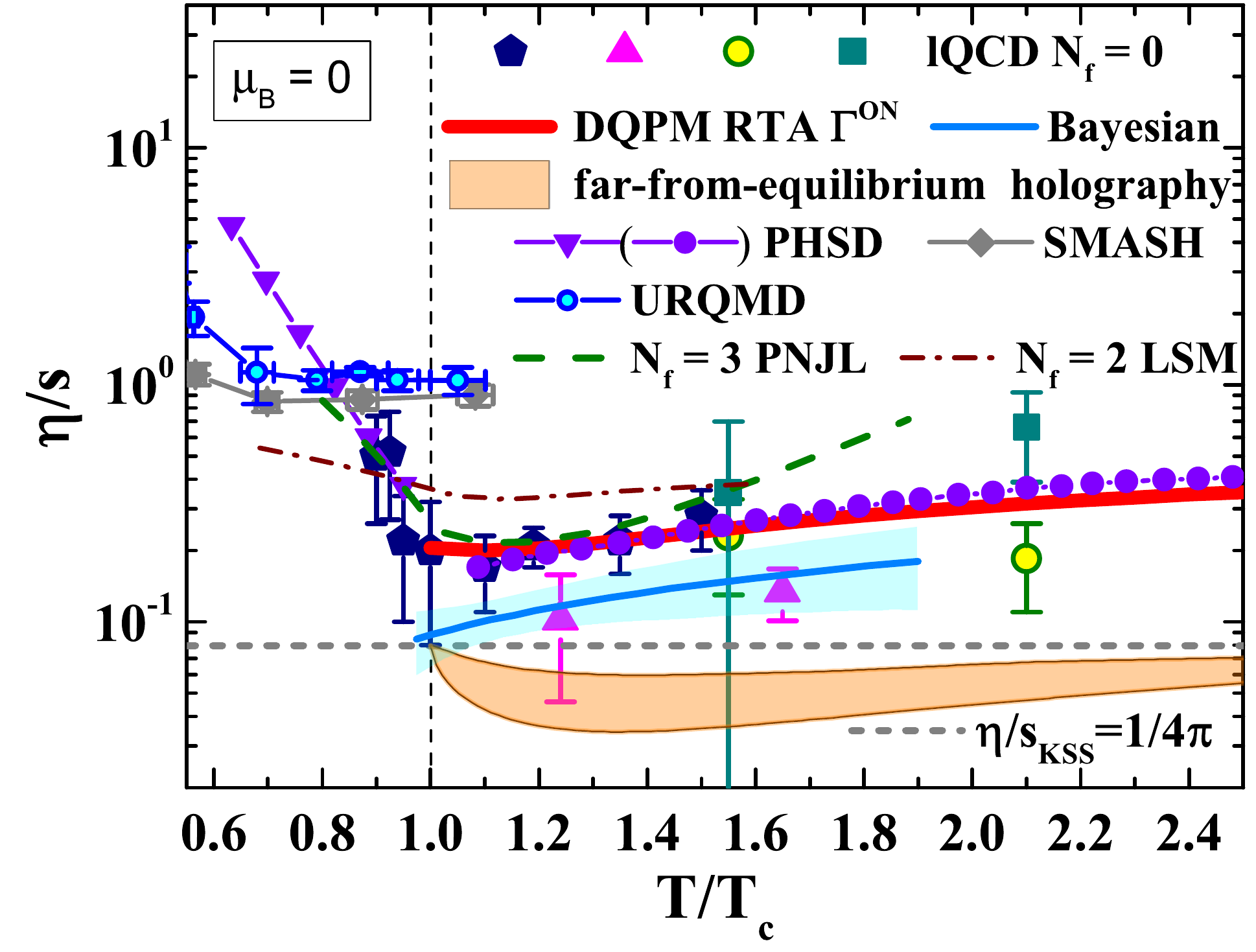}
 \includegraphics[scale=0.4]{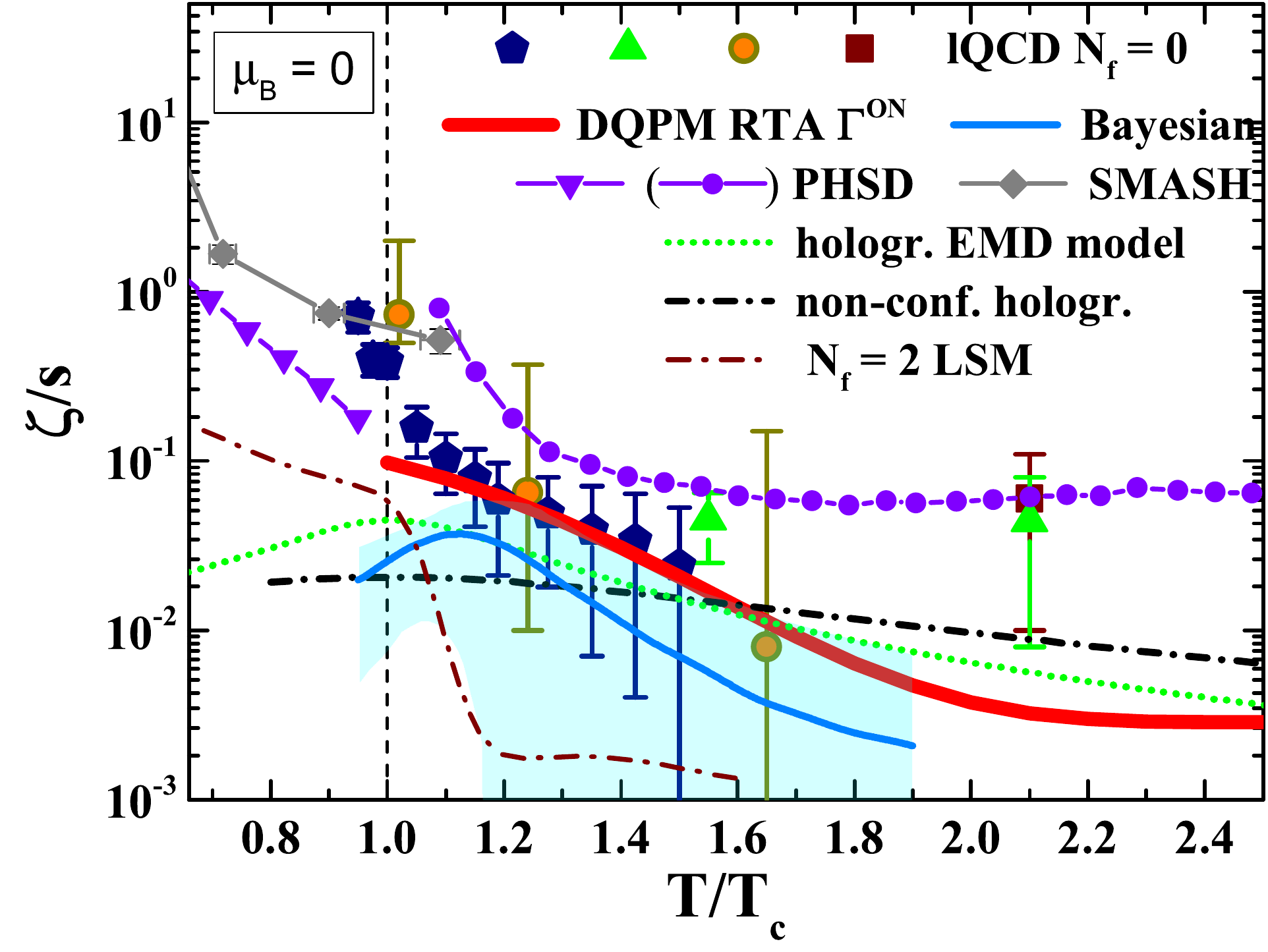}
\caption{Specific shear (left) and bulk (right) viscosities as a function of the scaled temperature $T/T_c$ at $\mu_B=0$. The DQPM results - obtained by the RTA approach with the interaction rate \cite{Soloveva:2019xph} - are presented by the red solid lines while  the dashed gray line demonstrates the Kovtun–Son–Starinets bound $(\eta/s)_{KSS}=  1/(4\pi)$ \cite{Kovtun:2004de}. The symbols corresponds to the lQCD results for pure SU(3) gauge theory (dark blue pentagons) \cite{Astrakhantsev:2017nrs,Astrakhantsev:2018oue}, (light green triangles and dark red squares) \cite{Nakamura:2004sy}, (orange circles) \cite{Meyer:2007dy} , (violet triangles)
\cite{Nakamura:2004sy}, (dark cyan squares) \cite{Meyer:2007ic}. The solid blue lines show the results from a Bayesian analysis of experimental heavy-ion data \cite{Bernhard:2019bmu}. The light green dotted line corresponds to the estimates from the Einstein-Maxwell-Dilaton (EMD) holographic model \cite{Rougemont:2017tlu}. The black dashed-dotted line depicts the predictions from the non-conformal holographic model \cite{Attems:2016ugt}. The orange area corresponds to the far-from-equilibrium time-dependent value of $\eta/s$ in a holographic model ( in a region of $t_{avg} = 0.24 - 0.59 fm$) \cite{Wondrak:2020tzt}. The dark red dashed-dotted line corresponds to the RTA estimations for the $N_f=2$ linear sigma model \cite{Heffernan:2020zcf}.
The dark green dashed line presents RTA estimations for the $N_f=3$ PNJL model \cite{Soloveva:2020hpr}.
For the hadronic phase $T<T_c = 0.158$ GeV we show evaluations from transport models: URQMD (blue line with open circles)
\cite{Demir:2008tr}, SMASH (grey solid line with rhombuses) \cite{Rose:2017bjz,Rose:2020lfc}.
The PHSD results for the hadronic phase are shown by the violet solid line with triangles
while for the partonic phase given by violet solid line with  circles \cite{Ozvenchuk:2012kh}. }
\label{fig:viscosities}
\end{figure}

The DQPM results of viscosities for the partonic phase are shown in Fig. \ref{fig:viscosities} (red lines) in comparison with the lQCD data, estimates from non-conformal holographic models \cite{Attems:2016ugt,Rougemont:2017tlu}, predictions from the Baesyan analysis of the experimental heavy-ion data \cite{Bernhard:2019bmu}, and the far-from-equilibrium time-dependent value of $\eta/s$ in a holographic model ( in a region of $t_{avg} = 0.24 - 0.59 fm$) \cite{Wondrak:2020tzt}.
Furthermore, the ratio of the shear viscosity $\eta$ over entropy density $s$, i.e.  $\eta/s$, as
evaluated within the relaxation time approximation (RTA) from the collisional widths, agrees
well with those calculated on the basis of the Kubo formalism, as well as with
lQCD calculations for pure SU(3) gauge theory \cite{Astrakhantsev:2017nrs,Astrakhantsev:2018oue,Nakamura:2004sy,Meyer:2007dy}
for $\mu_B =0$.
Moreover, we show the results for the hadronic phase from the transport models  URQMD (blue line with open circles)  
\cite{Demir:2008tr} and SMASH (grey solid line with rhombuses) \cite{Rose:2017bjz,Rose:2020lfc}.
The PHSD results for the hadronic phase are shown by the violet solid line with triangles and for the partonic phase by violet solid line with circles \cite{Ozvenchuk:2012kh}. 
The specific shear viscosity for the partonic phase from the PHSD is in a good agreement with the estimates from the DQPM, while in case of the bulk viscosity the previous PHSD result is higher than the actual DQPM calculations. This is related to the fact that in the early PHSD
study in Ref. \cite{Ozvenchuk:2012kh} the QGP phase has been realized with a DQPM  parametrization with slightly different quasiparticle  properties (masses and widths) that have been fitted to  the entropy density of earlier lQCD data \cite{Borsanyi:2010cj}. Also the present DQPM partonic cross sections are evaluated from leading order scattering diagrams and depend on
$T, \mu_B, \sqrt{s}$ and the scattering angel while the previous cross sections in Ref. \cite{Ozvenchuk:2012kh} were only depending
on the energy density, i.e. on temperature. The differences show up more pronounced
in the bulk viscosity than in the shear viscosity due to the terms related to the mass derivative and speed of sound $c_s^2$ in the expression for the bulk viscosity which are sensitive to the actual form of the temperature dependence of the masses
and cross sections.

The estimates of the specific shear viscosity in the  hadronic phase decrease when approaching the phase transition; meanwhile one can see a difference in the vicinity of the phase transition in the $T$-dependence for the transport models, which can be understood as a difference in the methods for the evaluation of the viscosities as well as the difference between the model description. Additionally, we show the RTA estimations from the $N_f=2$ linear sigma model (for vacuum sigma mass $m_{0}= 600$ MeV) \cite{Heffernan:2020zcf} and the $N_f=3$ PNJL model \cite{Soloveva:2020hpr}.

%----------------------------muB=0 viscosity compilation------------------------------------------------
%-------------------------------------------------------------------------------------------------------
%----------------------------muB=0 sigma compilation------------------------------------------------
\begin{figure}[!ht]
\includegraphics[scale=0.4]{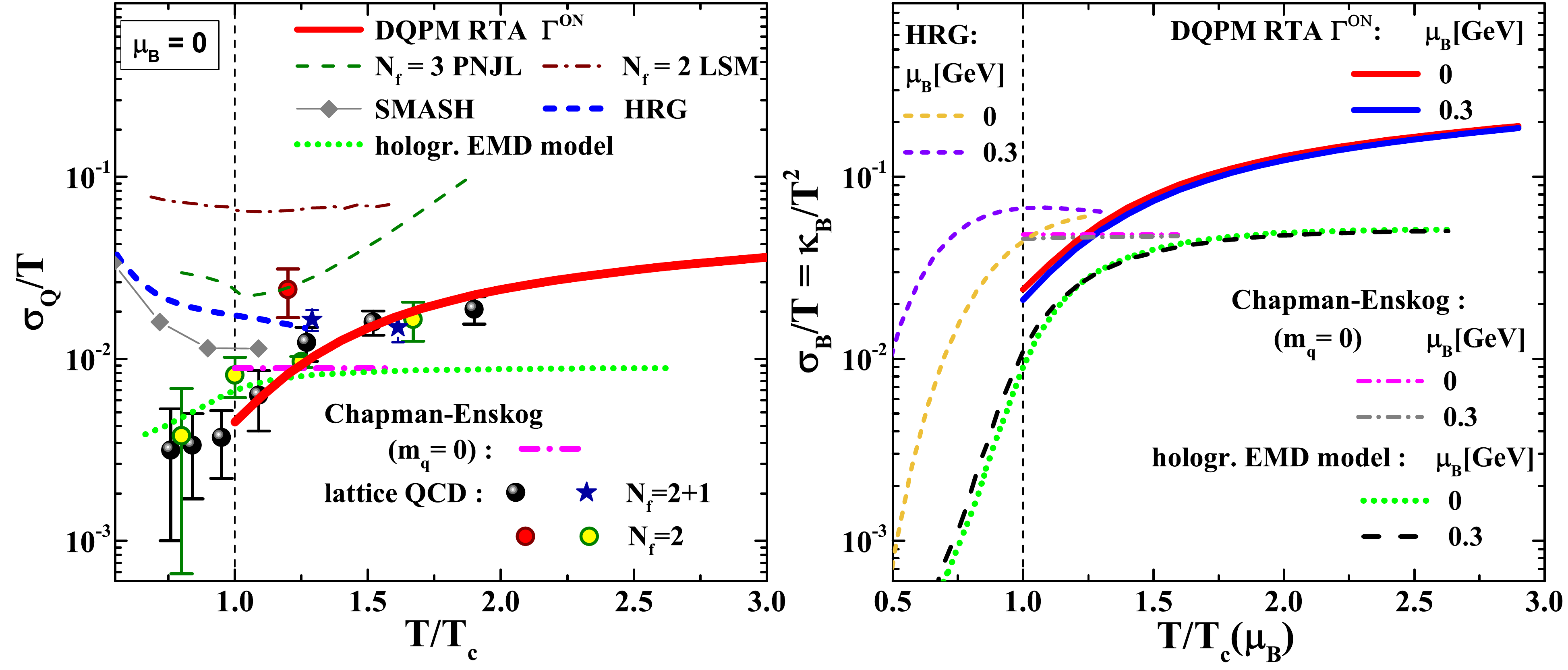}
\caption{Scaled electric (Left) and baryon (Right) conductivities as a function of scaled temperature $T/T_c$ at $\mu_B=0$. The DQPM results obtained by the RTA approach with the interaction rate \cite{Soloveva:2019xph} are presented by the red solid lines. The symbols corresponds to the lQCD results for $N_f=2:$ red circle-shaped point \cite{Brandt:2012jc}, yellow circle-shaped points \cite{Brandt:2015aqk}, and for $N_f=2+1:$ black spheres \cite{Aarts:2014nba} and blue stars \cite{Astrakhantsev:2019zkr}. Blue, violet and orange dashed lines show the estimations for HRG obtained within the Chapman-Enskog method \cite{Greif:2017byw}. The light green dotted line and dashed black line correspond to the estimations from the Einstein-Maxwell-Dilaton (EMD) holographic model \cite{Rougemont:2017tlu}. The grey solid line with rhombuses corresponds to the estimations from the SMASH transport model \cite{Rose:2020sjv}. The dark red dashed-dotted line corresponds to the RTA estimations for $N_f=2$ linear sigma model \cite{Heffernan:2020zcf}. The dark green dashed line presents RTA estimations for $N_f=3$ PNJL model \cite{Soloveva:2020hpr}.}
\label{fig:sigmas}
\end{figure}
%-------------------------------------------------------------------------------------------------------
%
Furthermore,  the electric conductivity and baryon conductivity of QGP matter, produced in HICs, are of fundamental importance.
It is known that the electric conductivity determines the soft photon spectra
\cite{Turbide:2003si,Akamatsu:2011mw,Linnyk:2015rco} and is directly related to their emission rate
\cite{Yin:2013kya}.
Moreover, the electric conductivity affects the generation and  evolution of electromagnetic fields produced in HICs \cite{Tuchin:2013apa,Inghirami:2019mkc,Denicol:2019iyh,Oliva:2020mfr}.
The DQPM results for the electric (left) and baryon conductivities  (right) are shown in Fig. \ref{fig:sigmas} as a function of the scaled temperature $T/T_c$.
We find a good agreement with the lQCD calculations for $\sigma_Q/T$ for $N_f=2:$ red circle-shaped point \cite{Brandt:2012jc}, yellow circle-shaped points \cite{Brandt:2015aqk}, and for $N_f=2+1:$ black spheres \cite{Aarts:2014nba} and blue stars \cite{Astrakhantsev:2019zkr}. The scaled electric and baryon conductivity have a similar temperature dependence: the ratios rise quadratically with temperature above $T_c$ which is due to the increasing quark density with temperature.
We compare also to the estimate from the Chapman-Enskog method using cross-sections for massless quarks and gluons as in Ref.~\cite{Greif:2017byw}, which are fixed  in order to describe the Kovtun-Son-Starinets bound for the shear viscosity to entropy density ratio $(\eta/s)_{KSS}=1/(4 \pi)$ \cite{Kovtun:2004de}, leading to $\sigma_{tot}\approx 0.72/T^2$. We also find a good agreement in the vicinity of $T_c$. Furthermore, there are  calculations for $\sigma_Q/T$ from the non-conformal holographic model \cite{Rougemont:2017tlu}, which are close to our results in the vicinity of the transition $T_c-1.5T_c$, however, the  temperature dependence of the ratio differs and the values at high temperatures are lower than the DQPM predictions.
There are also predictions for the electric conductivity by solving the relativistic transport equations for partons in a box with periodic boundary conditions in the presence of an external electric field as in Refs.~\cite{Cassing:2013iz,Puglisi:2014sha}.

The baryon conductivity is expected to be more sensitive to the net baryon density of the system.  Fig. \ref{fig:sigmas} (right) shows the actual results for the baryon diffusion coefficient in the range of temperature and non-zero baryon chemical potential $\mu_B = 0.3$ GeV as well as $\mu_B = 0$. We compare the DQPM results to the estimates from the non-conformal holographic model \cite{Rougemont:2017tlu} and the estimates for the hadron gas within the Chapman-Enskog method \cite{Greif:2017byw}. In the vicinity of $T_c$ the DQPM values for the diffusion coefficient are in agreement with the calculations within the Chapman-Enskog first-order approximation using cross-sections for massless quarks and gluons in Ref.~\cite{Greif:2017byw}. However, for higher temperatures the ratio $\sigma_B^{RTA}/T$ grows with temperature in the DQPM while the Chapman-Enskog results stay approximately constant $\sigma_B^{CE}/T^\sim0.048$ for all temperatures.  Recently, estimates for the full diffusion coefficient matrix of the QGP have shown rather small difference between the Chapman-Enskog method in comparison to the RTA method \cite{Fotakis:2021diq}, where the parton microscopic properties are described by the DQPM.
%----

\begin{figure}[!ht]
\begin{center}
 \includegraphics[scale=0.27]{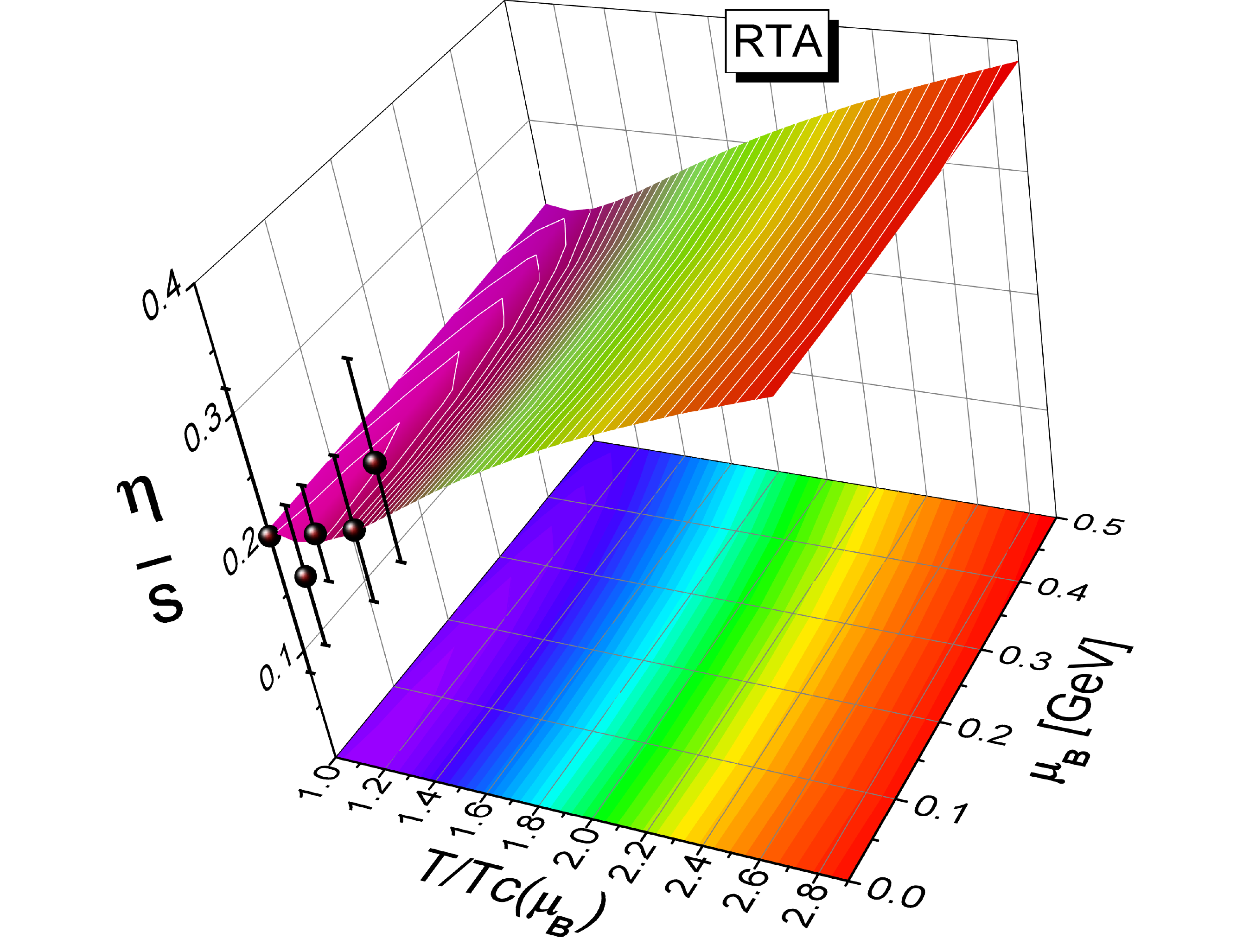}
 \includegraphics[scale=0.27]{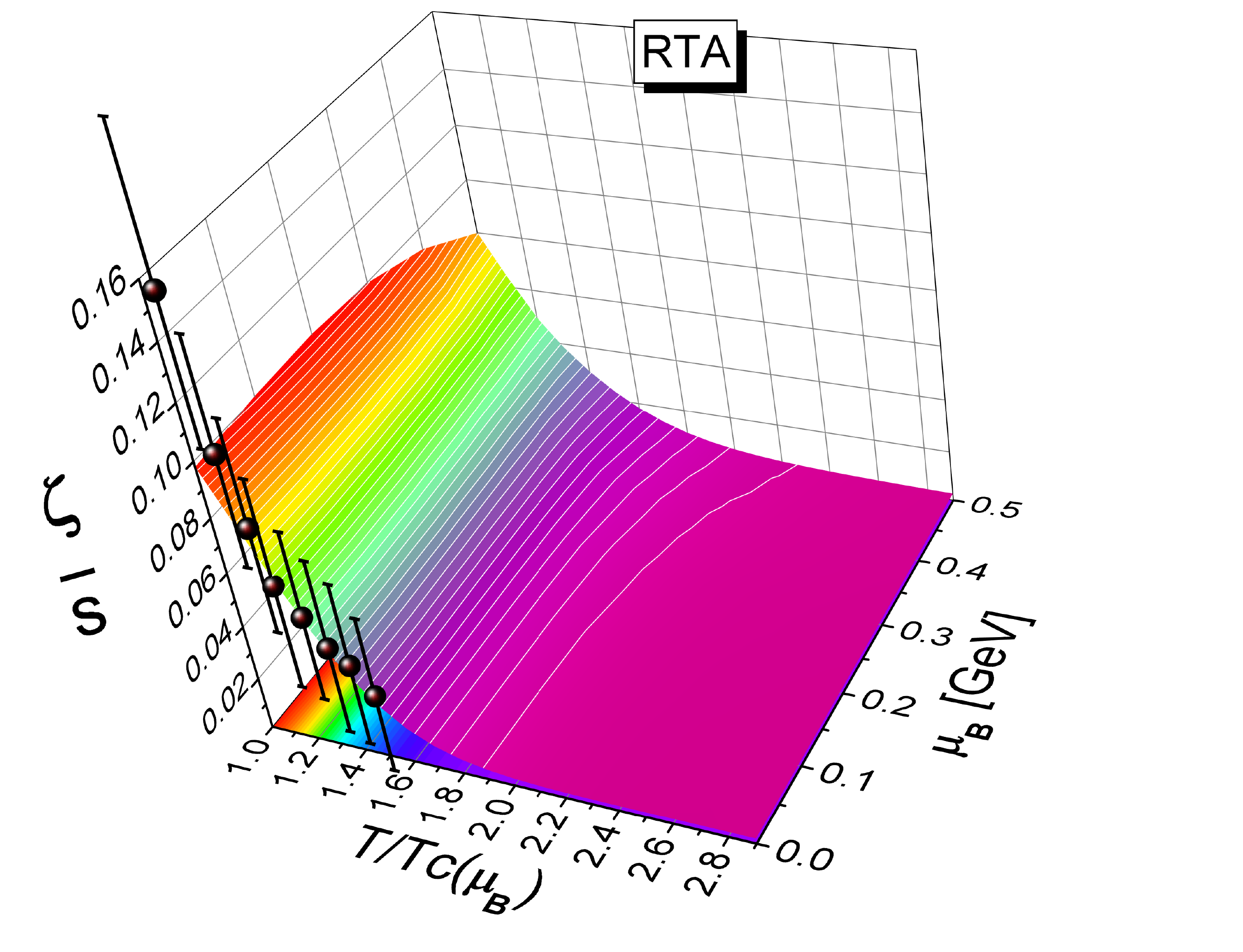}
 \includegraphics[scale=0.27]{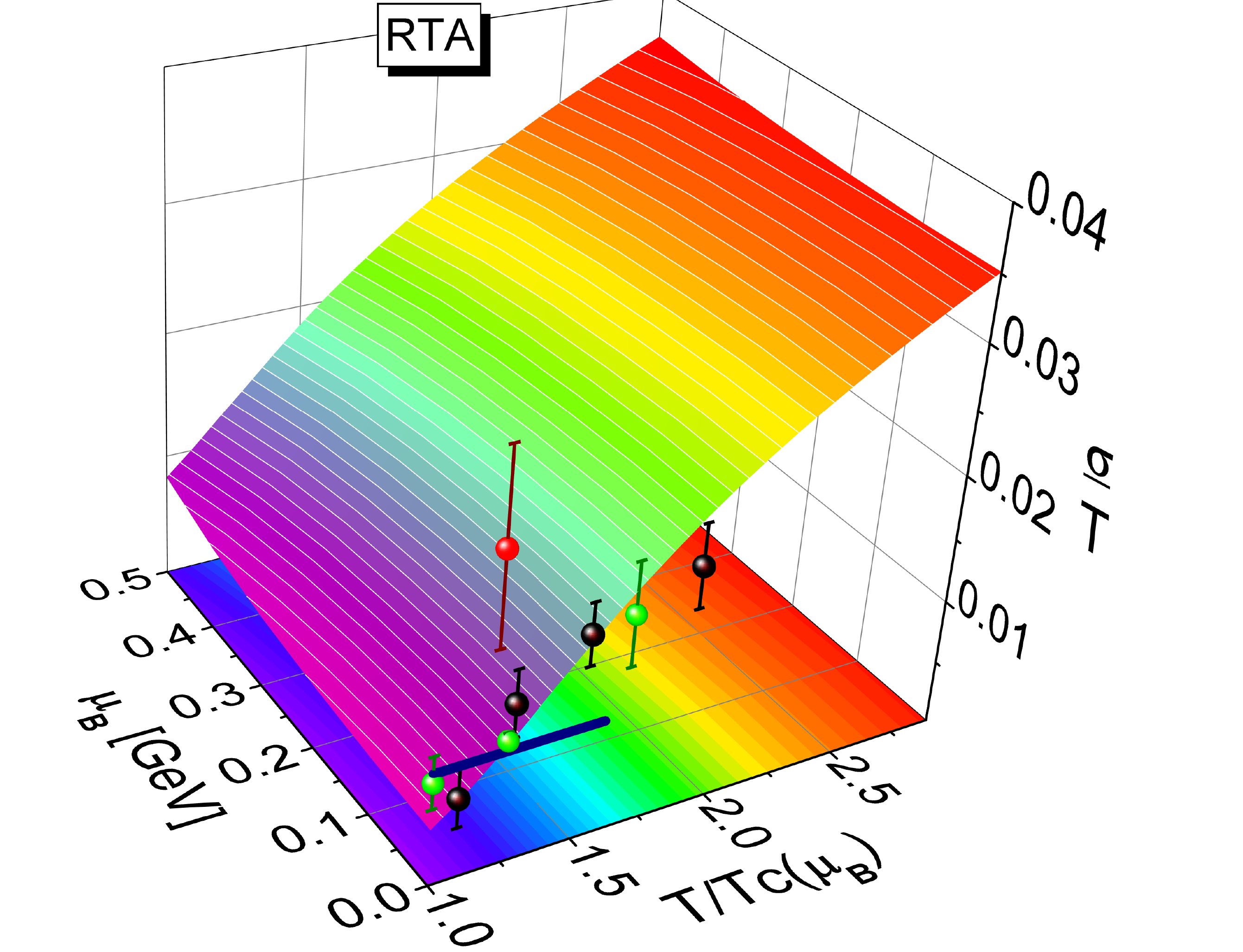}
 \includegraphics[scale=0.27]{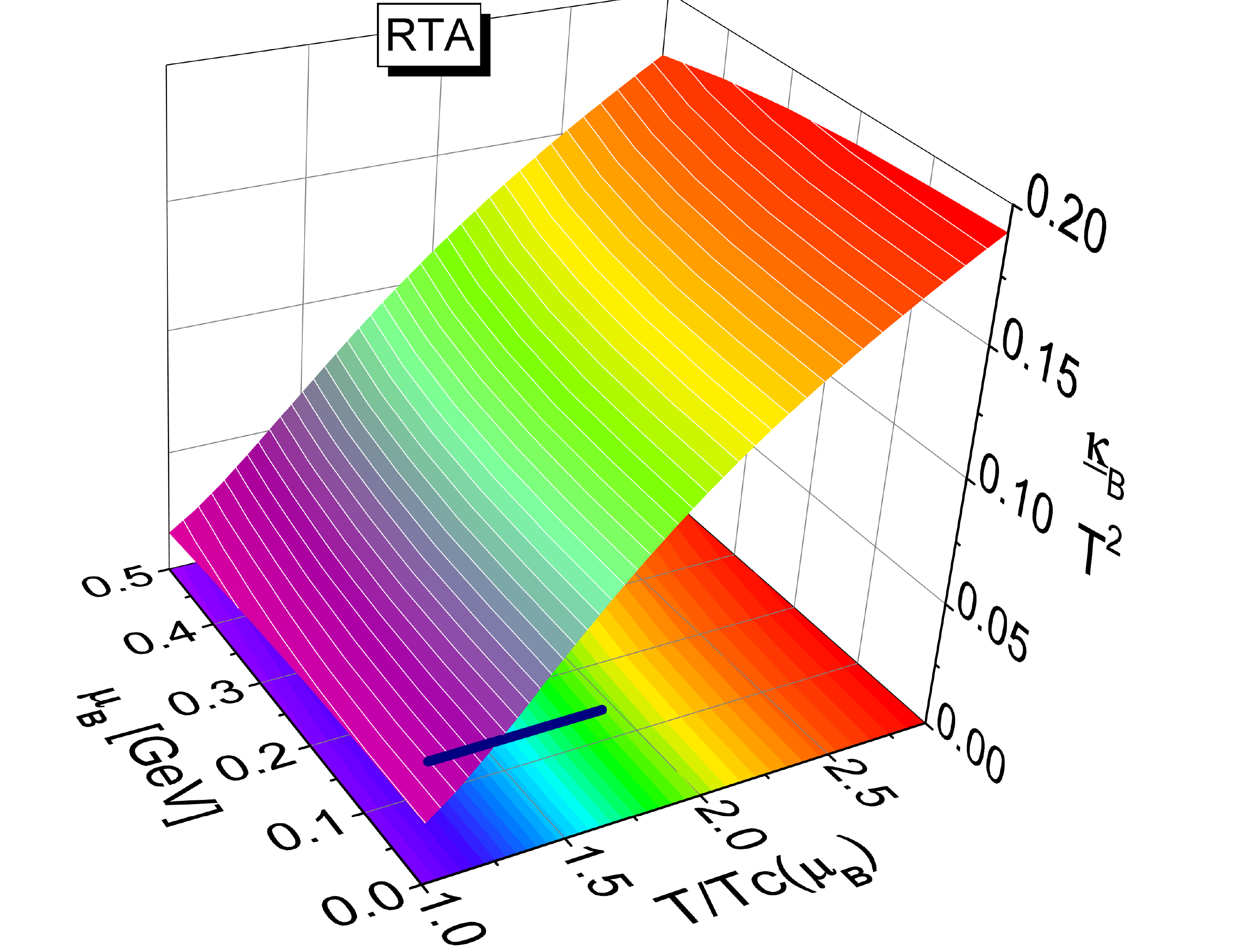}
\end{center}
\caption{The DQPM results for the ratios of shear viscosity to
entropy density $\eta/s$, bulk viscosity to entropy density
$\zeta/s$, electric conductivity to temperature
$\sigma_0/T$, the baryon-diffusion coefficient to the
temperature squared $\kappa_B/T^2$ for non-zero
$\mu_B$ as a function of the scaled temperature
$T/T_c(\mu_B)$ and the baryon chemical potential $\mu_B$ from
Ref.\ \cite{Soloveva:2019xph}.
The solid blue lines correspond to the results from the first-order
Chapman-Enskog approximation taken from Ref.\ \cite{Greif:2017byw}.
The lQCD calculations for $\mu_B=0$ are taken from Refs.\
\cite{Astrakhantsev:2017nrs,Astrakhantsev:2018oue,Brandt:2012jc,Brandt:2015aqk,Amato:2013naa,Aarts:2014nba}.
          }
\label{fig:etas}
\end{figure}

The DQPM results for non-zero baryon chemical potential, obtained within the relaxation-time approximation, are presented in Fig.\ \ref{fig:etas}. Here we
compare the DQPM calculations for $\sigma_Q/T$ and $\sigma_B/T\equiv\kappa_B/T^2$ to
the results from the first-order Chapman-Enskog approximation for a
simplified pQCD medium at $\mu_B=0$ taken from Ref.\
\cite{Greif:2017byw} and find a reasonable agreement.

%\end{document}

We find only a rather weak dependence of the transport coefficients on
$\mu_B$. The shear and bulk viscosities, and electric
conductivities of the QGP increase slightly with
increasing $\mu_B$ while the baryon-diffusion coefficient
decreases. The latter has important consequences for the transport
results for HIC observables, e.g., the baryon diffusion might enhance the
difference between proton and antiproton elliptic flow $v_2(p_T)$ and
mean transverse momenta.

\section{Dynamics of the QCD at finite chemical potential}
To study the evolution of the QCD medium produced in HICs one can rely on  microscopic transport approaches, which provide the full time evolution of the system. Nevertheless, the evolution of the deconfined QCD phase has been successfully described within  hydrodynamic simulations and hybrid methods, too. The parton-hadron-string dynamics (PHSD) approach is an off-shell transport approach based on Kadanoff-Baym equations in first-order gradient expansion  which allows for simulations of both the hadronic and the partonic phases. Whereas the hadronic part is essentially equivalent to the conventional HSD approach \cite{Cassing:1999es},  the partonic dynamics is based on the dynamical quasi-particle model (DQPM) described above. The phase transition from the partonic to hadronic degrees-of-freedom (for light quarks/antiquarks) is described by the dynamical hadronization, where covariant transition rates for the fusion of quark-antiquark pairs to mesonic resonances (or three quarks (antiquarks) to baryonic states) are used \cite{Cassing:2009vt,Bratkovskaya:2011wp}.
We mention that an out-of equilibrium study on the $\mu_B$ dependence of the QGP - created in HICs - has been performed within the PHSD transport approach,
extended in the partonic sector by explicitly calculating the total and
differential partonic scattering cross-sections based on the DQPM and
evaluated at the actual temperature $T$ and baryon chemical potential
$\mu_B$ in each individual space-time cell where partonic
scattering takes place \cite{Moreau:2019vhw,Soloveva:2020ozg}.

We found only a small influence of a $\mu_B$ dependence of
the QGP in heavy-ion observables, since when decreasing the collision
energy and thus increasing $\mu_B$, the QGP volume decreases
rapidly and the hadronic phase becomes dominant \cite{Moreau:2019vhw,Soloveva:2020ozg,Moreau:2021clr}.
Therefore even for low collision energies the influence of the QGP matter is washed out. Nevertheless, observables for strange hadrons -- kaons and especially anti-strange
hyperons -- as well as for antiprotons show a more pronounced effect of the
$\mu_B$ dependence of QGP interactions. This gives an experimental hint
for the search of finite-$\mu_B$ traces of the QGP for experiments
at the future FAIR and NICA accelerators, even if it will be a very
challenging experimental task.

\section{Conclusions and Outlook}
We have presented recent results on the transport properties of the QCD at finite chemical potential. Starting with the description of the microscopic properties of the parton degrees of freedom we have discussed the transport coefficients of the partonic phase obtained within the dynamical quasi-particle model (DQPM) and compared the results at vanishing chemical potential with numerous predictions from the literature.  We have found a good agreement with the available lQCD calculations for transport coefficients at $\mu_B$=0. Furthermore, we have shown that the presence of a non-zero chemical potential can affect the transport coefficients already in the region of moderate  baryon chemical potential $\mu_B \leq 500$ GeV, where a cross-over phase transition
likely takes place.

\section*{Acknowledgements}
The authors acknowledge inspiring discussions with J\"org Aichelin, Wolfgang Cassing, Lucia Oliva and Taesoo Song.
O.S. acknowledges support from the Helmholtz Graduate School
for Heavy Ion research.
P.M. acknowledges support by the U.S. D.O.E. under Grant No. DE-FG02-05ER41367.
Furthermore, we acknowledge support by the Deutsche Forschungsgemeinschaft
(DFG, German Research Foundation):  grant
CRC-TR 211 ’Strong-interaction matter under extreme conditions'  -  Project number  315477589  -  TRR  211;
by the European Union’s Horizon 2020 research and innovation program under grant agreement
No 824093 (STRONG-2020) and by the COST Action THOR, CA15213.
The computational resources have been provided by
the LOEWE-Center for Scientific Computing and the "Green Cube" at GSI, Darmstadt.

\section*{References}

%\bibliography{Biblio-brat}
\bibliography{Biblio_brat}

\providecommand{\newblock}{}
\begin{thebibliography}{10}
\expandafter\ifx\csname url\endcsname\relax
  \def\url#1{{\tt #1}}\fi
\expandafter\ifx\csname urlprefix\endcsname\relax\def\urlprefix{URL }\fi
\providecommand{\eprint}[2][]{\url{#2}}

\bibitem{Arsene:2004fa}
Arsene I {\em et~al.\/} (BRAHMS) 2005 {\em Nucl. Phys. A\/} {\bf 757} 1--27
  (\textit{Preprint} \eprint{nucl-ex/0410020})

\bibitem{Adams:2005dq}
Adams J {\em et~al.\/} (STAR) 2005 {\em Nucl. Phys. A\/} {\bf 757} 102--183
  (\textit{Preprint} \eprint{nucl-ex/0501009})

\bibitem{Shuryak:1978ij}
Shuryak E~V 1978 {\em Phys. Lett. B\/} {\bf 78} 150

\bibitem{Gyulassy:2004zy}
Gyulassy M and McLerran L 2005 {\em Nucl. Phys. A\/} {\bf 750} 30--63
  (\textit{Preprint} \eprint{nucl-th/0405013})

\bibitem{Odyniec:2019kfh}
Odyniec G (STAR) 2019 {\em PoS\/} {\bf CORFU2018} 151

\bibitem{Senger:2020fvj}
Senger P 2020 {\em JPS Conf. Proc.\/} {\bf 32} 010092

\bibitem{Sissakian:2009zza}
Sissakian A and Sorin A (NICA) 2009 {\em J. Phys. G\/} {\bf 36} 064069

\bibitem{Peshier:2005pp}
Peshier A and Cassing W 2005 {\em Phys. Rev. Lett.\/} {\bf 94} 172301
  (\textit{Preprint} \eprint{hep-ph/0502138})

\bibitem{Cassing:2007nb}
Cassing W 2007 {\em Nucl. Phys. A\/} {\bf 795} 70--97 (\textit{Preprint}
  \eprint{0707.3033})

\bibitem{Cassing:2007yg}
Cassing W 2007 {\em Nucl. Phys. A\/} {\bf 791} 365--381 (\textit{Preprint}
  \eprint{0704.1410})

\bibitem{Borsanyi:2012cr}
Borsanyi S, Endrodi G, Fodor Z, Katz S, Krieg S, Ratti C and Szabo K 2012 {\em
  JHEP\/} {\bf 08} 053 (\textit{Preprint} \eprint{1204.6710})

\bibitem{Borsanyi:2013bia}
Borsanyi S, Fodor Z, Hoelbling C, Katz S~D, Krieg S and Szabo K~K 2014 {\em
  Phys. Lett. B\/} {\bf 730} 99--104 (\textit{Preprint} \eprint{1309.5258})

\bibitem{Cassing:2008nn}
Cassing W 2009 {\em Eur. Phys. J. ST\/} {\bf 168} 3--87 (\textit{Preprint}
  \eprint{0808.0715})

\bibitem{Berrehrah:2016vzw}
Berrehrah H, Bratkovskaya E, Steinert T and Cassing W 2016 {\em Int. J. Mod.
  Phys. E\/} {\bf 25} 1642003 (\textit{Preprint} \eprint{1605.02371})

\bibitem{Cea:2014xva}
Cea P, Cosmai L and Papa A 2014 {\em Phys. Rev. D\/} {\bf 89} 074512
  (\textit{Preprint} \eprint{1403.0821})

\bibitem{Guenther:2020vqg}
Guenther J~N 2020  (\textit{Preprint} \eprint{2010.15503})

\bibitem{Borsanyi:2020fev}
Borsanyi S, Fodor Z, Guenther J~N, Kara R, Katz S~D, Parotto P, Pasztor A,
  Ratti C and Szabo K~K 2020 {\em Phys. Rev. Lett.\/} {\bf 125} 052001
  (\textit{Preprint} \eprint{2002.02821})

\bibitem{Bazavov:2018mes}
Bazavov A {\em et~al.\/} (HotQCD) 2019 {\em Phys. Lett. B\/} {\bf 795} 15--21
  (\textit{Preprint} \eprint{1812.08235})

\bibitem{Bellac:2011kqa}
Bellac M~L 2011 {\em {Thermal Field Theory}\/} Cambridge Monographs on
  Mathematical Physics (Cambridge University Press) ISBN 978-0-511-88506-8,
  978-0-521-65477-7

\bibitem{Linnyk:2015rco}
Linnyk O, Bratkovskaya E and Cassing W 2016 {\em Prog. Part. Nucl. Phys.\/}
  {\bf 87} 50--115 (\textit{Preprint} \eprint{1512.08126})

\bibitem{Baym}
Vanderheyden B and Baym G 1998 {\em J. Stat. Phys.\/} [J. Statist.
  Phys.93,843(1998)] (\textit{Preprint} \eprint{hep-ph/9803300})

\bibitem{Blaizot:2000fc}
Blaizot J~P, Iancu E and Rebhan A 2001 {\em Phys. Rev.\/} {\bf D63} 065003
  (\textit{Preprint} \eprint{hep-ph/0005003})

\bibitem{Soloveva:2019xph}
Soloveva O, Moreau P and Bratkovskaya E 2020 {\em Phys. Rev. C\/} {\bf 101}
  045203 (\textit{Preprint} \eprint{1911.08547})

\bibitem{Moreau:2019vhw}
Moreau P, Soloveva O, Oliva L, Song T, Cassing W and Bratkovskaya E 2019 {\em
  Phys. Rev. C\/} {\bf 100} 014911 (\textit{Preprint} \eprint{1903.10257})

\bibitem{Ryu:2015vwa}
Ryu S, Paquet J~F, Shen C, Denicol G~S, Schenke B, Jeon S and Gale C 2015 {\em
  Phys. Rev. Lett.\/} {\bf 115} 132301 (\textit{Preprint} \eprint{1502.01675})

\bibitem{Kubo:1957mj}
Kubo R 1957 {\em J. Phys. Soc. Jap.\/} {\bf 12} 570--586

\bibitem{Aarts:2002cc}
Aarts G and Martinez~Resco J~M 2002 {\em JHEP\/} {\bf 04} 053
  (\textit{Preprint} \eprint{hep-ph/0203177})

\bibitem{Ozvenchuk:2012kh}
Ozvenchuk V, Linnyk O, Gorenstein M~I, Bratkovskaya E~L and Cassing W 2013 {\em
  Phys. Rev. C\/} {\bf 87} 064903 (\textit{Preprint} \eprint{1212.5393})

\bibitem{Chakraborty:2010fr}
Chakraborty P and Kapusta J 2011 {\em Phys. Rev. C\/} {\bf 83} 014906
  (\textit{Preprint} \eprint{1006.0257})

\bibitem{Attems:2016ugt}
Attems M, Casalderrey-Solana J, Mateos D, Papadimitriou I, Santos-Oliv\'an D,
  Sopuerta C~F, Triana M and Zilh\~ao M 2016 {\em JHEP\/} {\bf 10} 155
  (\textit{Preprint} \eprint{1603.01254})

\bibitem{Rougemont:2017tlu}
Rougemont R, Critelli R, Noronha-Hostler J, Noronha J and Ratti C 2017 {\em
  Phys. Rev. D\/} {\bf 96} 014032 (\textit{Preprint} \eprint{1704.05558})

\bibitem{Bernhard:2019bmu}
Bernhard J~E, Moreland J~S and Bass S~A 2019 {\em Nature Phys.\/} {\bf 15}
  1113--1117

\bibitem{Wondrak:2020tzt}
Wondrak M~F, Kaminski M and Bleicher M 2020 {\em Phys. Lett. B\/} {\bf 811}
  135973 (\textit{Preprint} \eprint{2002.11730})

\bibitem{Astrakhantsev:2017nrs}
Astrakhantsev N, Braguta V and Kotov A 2017 {\em JHEP\/} {\bf 04} 101
  (\textit{Preprint} \eprint{1701.02266})

\bibitem{Astrakhantsev:2018oue}
Astrakhantsev N, Braguta V and Kotov A 2018 {\em Phys. Rev. D\/} {\bf 98}
  054515 (\textit{Preprint} \eprint{1804.02382})

\bibitem{Nakamura:2004sy}
Nakamura A and Sakai S 2005 {\em Phys. Rev. Lett.\/} {\bf 94} 072305
  (\textit{Preprint} \eprint{hep-lat/0406009})

\bibitem{Meyer:2007dy}
Meyer H~B 2008 {\em Phys. Rev. Lett.\/} {\bf 100} 162001 (\textit{Preprint}
  \eprint{0710.3717})

\bibitem{Demir:2008tr}
Demir N and Bass S~A 2009 {\em Phys. Rev. Lett.\/} {\bf 102} 172302
  (\textit{Preprint} \eprint{0812.2422})

\bibitem{Rose:2017bjz}
Rose J~B, Torres-Rincon J~M, Sch\"afer A, Oliinychenko D~R and Petersen H 2018
  {\em Phys. Rev. C\/} {\bf 97} 055204 (\textit{Preprint} \eprint{1709.03826})

\bibitem{Rose:2020lfc}
Rose J~B, Torres-Rincon J~M and Elfner H 2020 {\em J. Phys. G\/} {\bf 48}
  015005 (\textit{Preprint} \eprint{2005.03647})

\bibitem{Borsanyi:2010cj}
Borsanyi S, Endrodi G, Fodor Z, Jakovac A, Katz S~D, Krieg S, Ratti C and Szabo
  K~K 2010 {\em JHEP\/} {\bf 11} 077 (\textit{Preprint} \eprint{1007.2580})

\bibitem{Heffernan:2020zcf}
Heffernan M, Jeon S and Gale C 2020 {\em Phys. Rev. C\/} {\bf 102} 034906
  (\textit{Preprint} \eprint{2005.12793})

\bibitem{Soloveva:2020hpr}
Soloveva O, Fuseau D, Aichelin J and Bratkovskaya E 2020  (\textit{Preprint}
  \eprint{2011.03505})

\bibitem{Kovtun:2004de}
Kovtun P, Son D~T and Starinets A~O 2005 {\em Phys. Rev. Lett.\/} {\bf 94}
  111601 (\textit{Preprint} \eprint{hep-th/0405231})

\bibitem{Meyer:2007ic}
Meyer H~B 2007 {\em Phys. Rev. D\/} {\bf 76} 101701 (\textit{Preprint}
  \eprint{0704.1801})

\bibitem{Brandt:2012jc}
Brandt B~B, Francis A, Meyer H~B and Wittig H 2013 {\em JHEP\/} {\bf 03} 100
  (\textit{Preprint} \eprint{1212.4200})

\bibitem{Brandt:2015aqk}
Brandt B~B, Francis A, J\"ager B and Meyer H~B 2016 {\em Phys. Rev. D\/} {\bf
  93} 054510 (\textit{Preprint} \eprint{1512.07249})

\bibitem{Aarts:2014nba}
Aarts G, Allton C, Amato A, Giudice P, Hands S and Skullerud J~I 2015 {\em
  JHEP\/} {\bf 02} 186 (\textit{Preprint} \eprint{1412.6411})

\bibitem{Astrakhantsev:2019zkr}
Astrakhantsev N, Braguta V~V, D'Elia M, Kotov A~Y, Nikolaev A~A and Sanfilippo
  F 2020 {\em Phys. Rev. D\/} {\bf 102} 054516 (\textit{Preprint}
  \eprint{1910.08516})

\bibitem{Greif:2017byw}
Greif M, Fotakis J~A, Denicol G~S and Greiner C 2018 {\em Phys. Rev. Lett.\/}
  {\bf 120} 242301 (\textit{Preprint} \eprint{1711.08680})

\bibitem{Rose:2020sjv}
Rose J~B, Greif M, Hammelmann J, Fotakis J~A, Denicol G~S, Elfner H and Greiner
  C 2020 {\em Phys. Rev. D\/} {\bf 101} 114028 (\textit{Preprint}
  \eprint{2001.10606})

\bibitem{Turbide:2003si}
Turbide S, Rapp R and Gale C 2004 {\em Phys. Rev. C\/} {\bf 69} 014903
  (\textit{Preprint} \eprint{hep-ph/0308085})

\bibitem{Akamatsu:2011mw}
Akamatsu Y, Hamagaki H, Hatsuda T and Hirano T 2011 {\em J. Phys. G\/} {\bf 38}
  124184 (\textit{Preprint} \eprint{1106.5870})

\bibitem{Yin:2013kya}
Yin Y 2014 {\em Phys. Rev. C\/} {\bf 90} 044903 (\textit{Preprint}
  \eprint{1312.4434})

\bibitem{Tuchin:2013apa}
Tuchin K 2013 {\em Phys. Rev. C\/} {\bf 88} 024911 (\textit{Preprint}
  \eprint{1305.5806})

\bibitem{Inghirami:2019mkc}
Inghirami G, Mace M, Hirono Y, Del~Zanna L, Kharzeev D~E and Bleicher M 2019
  (\textit{Preprint} \eprint{1908.07605})

\bibitem{Denicol:2019iyh}
Denicol G~S, Moln\'ar E, Niemi H and Rischke D~H 2019 {\em Phys. Rev. D\/} {\bf
  99} 056017 (\textit{Preprint} \eprint{1902.01699})

\bibitem{Oliva:2020mfr}
Oliva L 2020 {\em Eur. Phys. J. A\/} {\bf 56} 255 (\textit{Preprint}
  \eprint{2007.00560})

\bibitem{Cassing:2013iz}
Cassing W, Linnyk O, Steinert T and Ozvenchuk V 2013 {\em Phys. Rev. Lett.\/}
  {\bf 110} 182301 (\textit{Preprint} \eprint{1302.0906})

\bibitem{Puglisi:2014sha}
Puglisi A, Plumari S and Greco V 2014 {\em Phys. Rev. D\/} {\bf 90} 114009
  (\textit{Preprint} \eprint{1408.7043})

\bibitem{Fotakis:2021diq}
Fotakis J~A, Soloveva O, Greiner C, Kaczmarek O and Bratkovskaya E 2021
  (\textit{Preprint} \eprint{2102.08140})

\bibitem{Amato:2013naa}
Amato A, Aarts G, Allton C, Giudice P, Hands S and Skullerud J~I 2013 {\em
  Phys. Rev. Lett.\/} {\bf 111} 172001 (\textit{Preprint} \eprint{1307.6763})

\bibitem{Cassing:1999es}
Cassing W and Bratkovskaya E 1999 {\em Phys. Rept.\/} {\bf 308} 65--233

\bibitem{Cassing:2009vt}
Cassing W and Bratkovskaya E 2009 {\em Nucl. Phys. A\/} {\bf 831} 215--242
  (\textit{Preprint} \eprint{0907.5331})

\bibitem{Bratkovskaya:2011wp}
Bratkovskaya E, Cassing W, Konchakovski V and Linnyk O 2011 {\em Nucl. Phys.
  A\/} {\bf 856} 162--182 (\textit{Preprint} \eprint{1101.5793})

\bibitem{Soloveva:2020ozg}
Soloveva O, Moreau P, Oliva L, Voronyuk V, Kireyeu V, Song T and Bratkovskaya E
  2020 {\em Particles\/} {\bf 3} 178--192 (\textit{Preprint}
  \eprint{2001.05395})

\bibitem{Moreau:2021clr}
Moreau P, Soloveva O, Grishmanovskii I, Voronyuk V, Oliva L, Song T, Kireyeu V,
  Coci G and Bratkovskaya E 2021 {Properties of the QGP created in heavy-ion
  collisions} {\em {9th International Workshop on Astronomy and Relativistic
  Astrophysics}\/} (\textit{Preprint} \eprint{2101.05688})

\end{thebibliography}
%\bibliography{iopart-num}
\end{document}